\documentclass[intlimits,twoside,a4paper]{article}
\usepackage[cp1251]{inputenc}
\usepackage[eqsecnum]{cmpj3}

\usepackage{bm}

\usepackage{multirow}
\usepackage[usenames]{color}
\usepackage{colortbl}

\issue{2020}{23}{4}{43001}
\doinumber{10.5488/CMP.23.43001}

\title[Preparation and study of the entanglement]%
{Preparation and study of the entanglement of the Schr\"odinger cat state on the ibmq-melbourne quantum computer%
}
\author[A.R. Kuzmak, V.M. Tkachuk]{A.R. Kuzmak,
        V.M. Tkachuk}
\address{
Department for Theoretical Physics, Ivan Franko National University of Lviv,\\
12 Drahomanov St., 79005 Lviv, Ukraine
}

\date{Received August 3, 2020, in final form October 3, 2020}

\begin{document}

\maketitle

\begin{abstract}
We study the entanglement between a certain qubit and the remaining system in the Schr\"odinger cat state
prepared on the ibmq-melbourne quantum computer. The protocol, which we use for this purpose, is based on the determination
of the mean value of spin corresponding to a certain qubit. We explore the dependence of the entanglement
on a parameter of the Schr\"odinger cat state which consists of different numbers of qubits.
In addition, we explore the entanglement of each qubit with the remaining system in the maximum entangled
Schr\"odinger cat state.

\end{abstract}

\section{Introduction}

The quantum computer is the best tool for the study of many-body quantum systems \cite{feynman1982,lloyd1996,loss1998,kane1998}.
As compared with classic computers, quantum computers provide an exponential supremacy in time to
simulate the evolution of large quantum systems. Recently, such a supremacy was shown on 53 qubits
of the Google quantum computer \cite{arute2019}. Using the quantum computer, the authors spent about 200
second, instead of 10\,000 years, to prepare and measure a quantum state on 53 qubits.
 This is because the classical simulation of specific classes of circuits past a certain depth on that many qubits is intractable.
However, the IBM's group argue that an ideal simulation of the same task can be performed on a classical system in 2.5 days and with far greater fidelity \cite{pednault2019}.

Another important property of quantum computers which allows us to efficiently simulate the behaviour of quantum systems
is quantum entanglement \cite{EPRP,horodecki2009}. It allows us to implement quantum
algorithms that have no analogue in the classical information \cite{nielsen2000}. To effectively
implement these algorithms, it is important to know the information on the value of the
entanglement of a system. Recently,  the negativity as a measure of entanglement
between certain pairs of qubits was measured on the IBM Q quantum devices \cite{wang2018,mooney2019}. The authors showed that the 16-qubit
\cite{wang2018} and 20-qubit \cite{mooney2019} quantum processors can be fully entangled. In our previous
paper \cite{kuzmak2020}, we proposed the protocol which allows one to determine the value of entanglement
between a certain qubit and the rest of a system. This protocol is based on the geometric measure of entanglement
defined by the mean value of the spin \cite{frydryszak2017}. It is important to stress that this is a bi-partite measure of entanglement rather than a multi-partite because it allows one to determine
the value of entanglement between two subsystems: a certain qubit and the rest qubits of state.
We tested this protocol on the 5-qubit ibmq-ourence
quntum computer.  Using this protocol, the value of entanglement of 2-, 3- and 4-qubit Schr\"odinger cat
and 3-qubit Werner states was determined. In addition, the entanglement of rank-2 two-qubit mixed states prepared on the ibmq-ourence quantum computer
was studied. Namely, we considered the entanglement of the mixed state that consists of two Bell states. In the present paper,
we apply the protocol to the states which consist of a larger number of qubits. We explore the behaviour of entaglement with an increasing number of qubits
in the Schr\"odinger cat state. For this purpose, we use the ibmq-melbourne quantum computer because it consists of 15 superconducting qubits.

The IBM has developed a cloud service called the IBM Q Experience \cite{IBMQExp,OpenQasm},
which permits a free access to various quantum devices based on superconducting qubits.
They consist of superconducting qubits controlled by four single-qubit basis gate and
a controlled-NOT gate. The controlled-NOT operators perform the interaction between qubits.
An arbitrary operator on these devices can be implemented by a combination of these gates.

\begin{figure}[!t]
\centering
\includegraphics[scale=0.600, angle=0.0, clip]{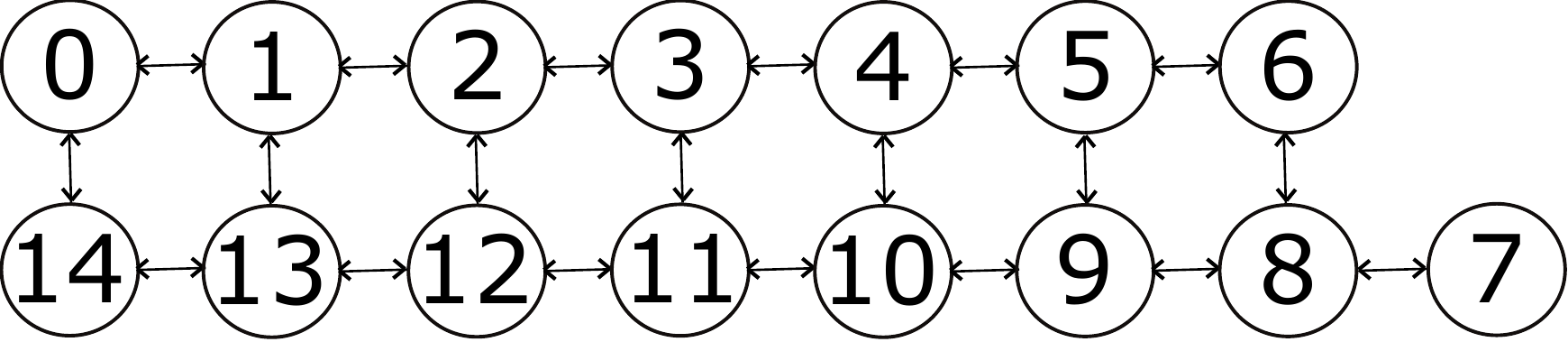}
\caption{Structure of the ibmq-melbourne quantum device.}
\label{ibmq_16_melbourne}
\end{figure}

In the present paper, using protocol \cite{kuzmak2020} to define the bi-partite geometric measure of entanglement, 
we prepare and study the entanglement
of the Schr\"odinger cat states prepared on the ibmq-melbourne quantum computer.
This state is used for the implementation of various quantum information processes because
it can have a maximum entanglement. We provide investigation for different
numbers of qubits in the state.
Therefore, we use the ibmq-melbourne quantum computer because it consists of a large number
of qubits and has a free access. This device has fifteen superconducting
qubits which are connected by the controlled-NOT gate in the way shown in figure~\ref{ibmq_16_melbourne}.
Bidirectionality of arrows means that each of the qubits can be both a control and a target.

\section{Preparation of the Schr\"odinger cat state on quantum computer}

\begin{figure}[!t]
\centering
\includegraphics[scale=0.7, angle=0.0, clip]{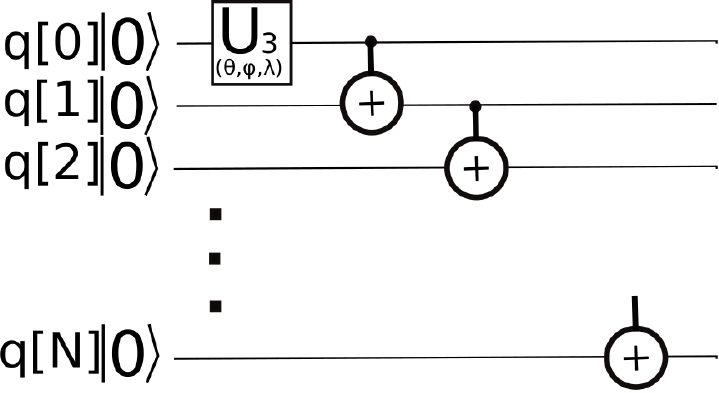}
\caption{Circuits for the preparation of a general Schr\"odinger cat state on a system of $N$ qubits.}
\label{HCatstate}
\end{figure}

The Schr\"odinger cat state which consists of $N$ qubits reads
\begin{eqnarray}
\vert\psi_{\textrm{cat}}\rangle =\cos\frac{\theta}{2}\vert 00\ldots 0\rangle + \re^{\ri\phi}\sin\frac{\theta}{2}\vert 11\ldots 1\rangle.
\label{Heisenbergstate}
\end{eqnarray}
This state can be prepared by applying  the single-qubit $U_3(\theta,\phi,\lambda)$ gate and a sequence
of controlled-NOT operators to the initial state
$\vert 00\ldots 0\rangle$ as it is shown in figure~\ref{HCatstate}. The gate $U_3(\theta,\phi,\lambda)$
is one of the four single-qubit basis operators of the IBM Q devices \cite{IBMQExp,OpenQasm}. In the basis
$\vert 0\rangle$, $\vert 1\rangle$ the matrix representation of $U_3(\theta,\phi,\lambda)$ gate can be expressed
as follows:
\begin{eqnarray}
U_3(\theta,\phi,\lambda)=\left( \begin{array}{ccccc}
\cos\frac{\theta}{2} & -\re^{\ri\lambda}\sin{\frac{\theta} {2}} \\[6pt]
\re^{\ri\phi}\sin{\frac{\theta} {2}} & \re^{\ri\left(\lambda+\phi\right)}\cos\frac{\theta}{2}
\end{array}\right).
\label{U3gate}
\end{eqnarray}
This gate has the effect of rotating a qubit in the initial state $\vert 0\rangle$ to an arbitrary one-qubit state
\begin{eqnarray}
U_3(\theta,\phi,\lambda)\vert 0\rangle =\cos\frac{\theta}{2}\vert 0\rangle+\re^{\ri\phi}\sin\frac{\theta}{2}\vert 1\rangle.
\label{arboneqstate}
\end{eqnarray}
A controlled-NOT gate is a basis multi-qubit gate of the IBM Q devices and it acts on a pair of qubits, with one acting as
`control' and the other as `target'. It provides the $\sigma^x$ Pauli operator on the target qubit whenever the control qubit
is in state $\vert 1\rangle$.

\begin{figure}[!t]
	\centering
\includegraphics[scale=0.49, angle=0.0, clip]{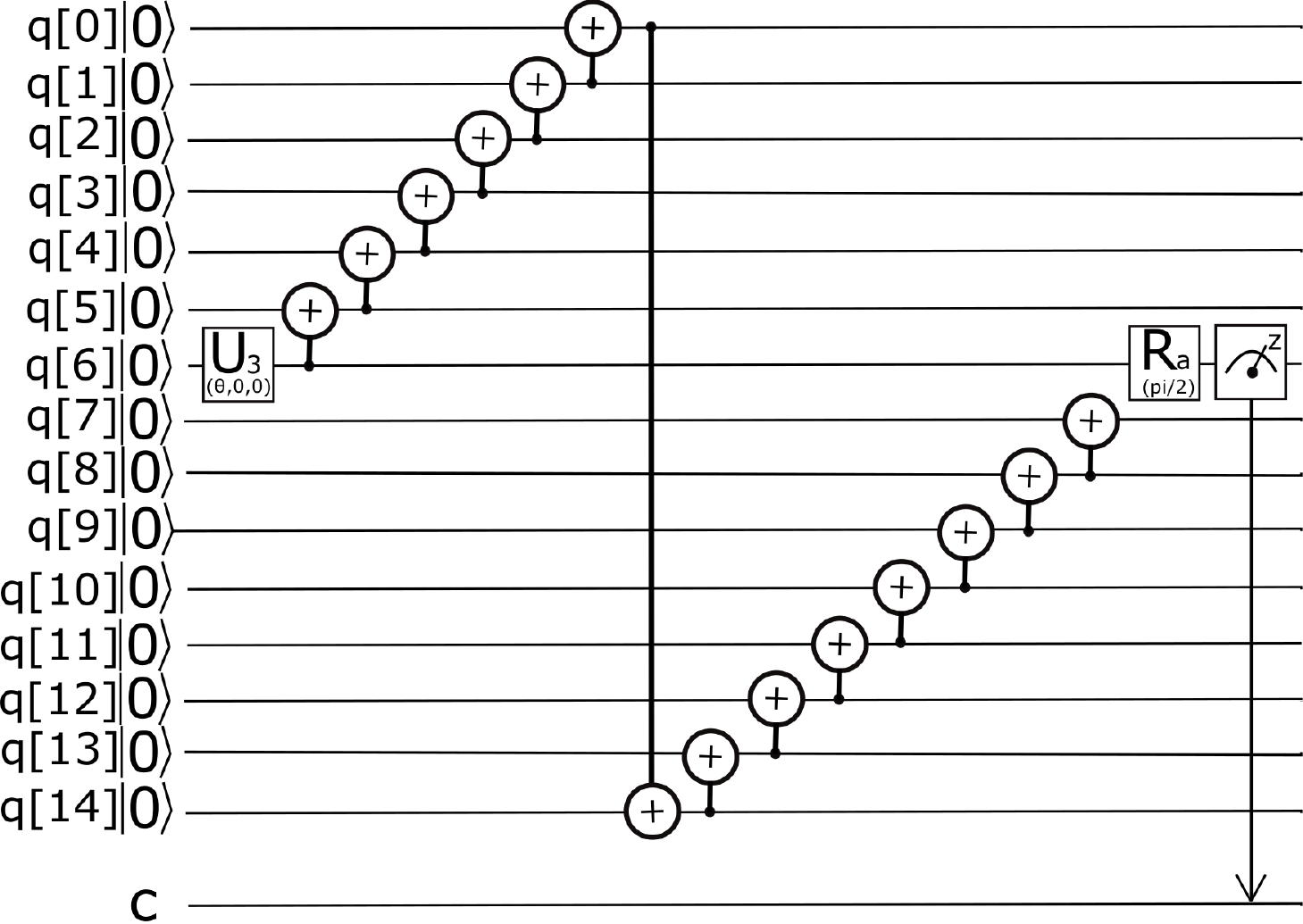}
	\caption{Circuit for the preparation and measurement of the 15-qubit Schr\"odinger cat state. Operators $R_a$ provide the rotation of the qubit state around the $a=x,y,z$ axis by the angle $\piup/2$.}
	\label{ShCatstate}
\end{figure}

According to the ibm-melbourne structure (see figure~\ref{ibmq_16_melbourne}), we prepare a 15-qubit Schr\"odinger cat state
in the way shown in figure~\ref{ShCatstate}. We use such a circuit because it allows us to perform the controlled-NOT gates only between
the neighbouring qubits on the ibmq-melbourne device. Using the protocol developed in paper \cite{kuzmak2020} we study the entanglement
of the Schr\"odinger cat state.

\section{Entanglement of the Schr\"odinger cat state}

We consider the protocol for measuring the entanglement between a certain qubit and the remaining system
on the quantum computer \cite{kuzmak2020}. This protocol is based on the geometric measure of entanglement
defined by the mean value of spin \cite{frydryszak2017}. Hence, any pure state of a quantum system can be expressed as
some unitary transforming $U$ of the state $\vert 00\ldots 0\rangle$
\begin{eqnarray}
\vert\psi\rangle=U\vert 00\ldots 0\rangle.
\label{systemstate}
\end{eqnarray}
Note that the state $\vert 00\ldots 0\rangle$ is the initial state of IBM Q devices. To identify
the entanglement between a certain qubit and the other qubits of the system we use the geometric measure
of entanglement which can be written as follows \cite{frydryszak2017,kuzmak2018,Golskyi2019}:
\begin{eqnarray}
E\left(\vert\psi\rangle\right)=\frac{1}{2}\left(1-\vert\langle\psi\vert{\bm \sigma}\vert\psi\rangle\vert\right),
\label{geommeasure}
\end{eqnarray}
where ${\bm \sigma}=\textbf{i}\sigma^x+\textbf{j}\sigma^y+\textbf{k}\sigma^z$ is the Pauli operator which corresponds to a certain qubit, and the
module of the mean value of this operator is determined by the expression
$\vert\langle\psi\vert{\bm \sigma}\vert\psi\rangle\vert=\sqrt{\langle\psi\vert{\bm \sigma}\vert\psi\rangle^2}$.
Since any two-level quantum system is described by the Pauli operator, expression (\ref{geommeasure}) can be used to determine the value
of entanglement of any quantum system that consists of a set of two-level systems. In the case considered in this paper,
we study the entanglement of the states prepared on a set of superconducting qubits. As it follows
from definition (\ref{geommeasure}), to obtain the value of entanglement of state $\vert\psi\rangle$, the mean values of
the Pauli operators $\langle\psi\vert\sigma^{\alpha}\vert\psi\rangle$ should be measured, where $\alpha=x,y,z$.
Note that quantum computers allow us to provide the measurements of qubits on the eigenstates
$\vert 0\rangle$, $\vert 1\rangle$ of $\sigma^z$ operator. Therefore, the mean values of the Pauli operators
should be expressed in the form \cite{kuzmak2020}
\begin{eqnarray}
&\langle\psi\vert\sigma^x\vert\psi\rangle=\vert\langle\tilde{\psi}^y\vert 0\rangle\vert^2-\vert\langle\tilde{\psi}^y\vert 1\rangle\vert^2,\quad
\langle\psi\vert\sigma^y\vert\psi\rangle=\vert\langle\tilde{\psi}^x\vert 0\rangle\vert^2-\vert\langle\tilde{\psi}^x\vert 1\rangle\vert^2,&\nonumber\\
&\langle\psi\vert\sigma^z\vert\psi\rangle=\vert\langle\psi\vert 0\rangle\vert^2-\vert\langle\psi\vert 1\rangle\vert^2,&
\label{mvpauli}
\end{eqnarray}
where $\vert\tilde{\psi}^y\rangle=\re^{\ri\frac{\piup}{4}\sigma^y}\vert\psi\rangle$, $\vert\tilde{\psi}^x\rangle=\re^{-\ri\frac{\piup}{4}\sigma^x}\vert\psi\rangle$.
Here, we use the fact that $\sigma^x=\re^{-\ri\frac{\piup}{4}\sigma^y}\sigma^z\re^{\ri\frac{\piup}{4}\sigma^y}$,
$\sigma^y=\re^{\ri\frac{\piup}{4}\sigma^x}\sigma^z\re^{-\ri\frac{\piup}{4}\sigma^x}$ and
$\sigma^z=\vert 0\rangle\langle 0\vert-\vert 1\rangle\langle 1\vert$.
As we can see, the mean value of the Pauli operators in state~$\vert\psi\rangle$ is defined by the probabilities
which define the result of measure of qubit on a basis vectors $\vert 0\rangle$, $\vert 1\rangle$.
In addition, before measuring the mean value of the $x$- and $y$-component of the spin, it should be rotated around the $y$- and $x$-axis
by the angle $\piup/2$. Due to the modules in expressions (\ref{mvpauli}), the directions of rotation are unimportant.
Thus, expressions (\ref{mvpauli}) allow one to express the mean values of the Pauli operators by the probabilities
which are measured experimentally on the quantum computer. Let us use this protocol to measure
the entanglement of the Schr\"odinger cat state (\ref{Heisenbergstate}) prepared on the ibmq-melbourne quantum computer.

To obtain the value of entanglement of any qubit with the remaining system in the Schr\"odinger cat state (\ref{Heisenbergstate}),
the mean values of the Pauli operators which correspond to a certain qubit should be calculated. In this state, they
have the following form:
\begin{eqnarray}
\langle\psi_{\textrm{cat}}\vert\sigma^z\vert\psi_{\textrm{cat}}\rangle=\cos\theta,\quad \langle\psi_{\textrm{cat}}\vert\sigma^x\vert\psi_{\textrm{cat}}\rangle=0,\quad \langle\psi_{\textrm{cat}}\vert\sigma^y\vert\psi_{\textrm{cat}}\rangle=0 .
\label{mvpshcats}
\end{eqnarray}
Using equation (\ref{geommeasure}) with mean values (\ref{mvpshcats}), the entanglement of any qubit with the remaining system in state (\ref{Heisenbergstate}) has the form
\begin{eqnarray}
E\left(\vert\psi_{\textrm{cat}}\rangle\right)=\frac{1}{2}\left(1-\vert\cos\theta\vert\right).
\label{geommeasurecat}
\end{eqnarray}

\begin{table}[!b]
	\vspace{-.3cm}
	\centering
	\caption{Calibration parameters of the ibmq-melbourne quantum device, archived 04 April 2020 from reference \cite{IBMQExp}.}
	\vspace{2ex}
	\scriptsize{
		\begin{tabular}{ c c c c c c c c c c c c c c c c }
			& \textbf{Q0} & \textbf{Q1} & \textbf{Q2} & \textbf{Q3} & \textbf{Q4} & \textbf{Q5} & \textbf{Q6} & \textbf{Q7} & \textbf{Q8} \\
			\\
			$T_1$, {\textmu}s & 69.0 & 63.2 & 55.9 & 71.0 & 69.0 & 18.8 & 89.2 & 46.0 & 40.8 \\
			\\
			$T_2$, {\textmu}s & 22.8 & 73.6 & 96.8 & 54.8 & 45.6 & 32.0 & 113.9 & 82.2 & 71.2 \\
			\\
			\textbf{Gate Error ($10^{-3}$)} & 0.61 &  1.36 & 1.76  & 0.66  & 1.72 & 3.31  & 0.83  & 1.34 & 0.63 \\
			\\
			\textbf{Readout Error ($10^{-2}$)} & 3.05 & 2.65 & 4.85 & 5.95 & 3.30 & 6.85 & 2.60 & 3.05 & 2.80 \\
			\\
			\multirow{1}{*}{\textbf{Multi-Qubit Gate (CX)}}& \textbf{0\_1} & \textbf{1\_2} & \textbf{2\_3} & \textbf{3\_4} & \textbf{4\_5} & \textbf{5\_6} & \textbf{6\_8} & \textbf{7\_8} & \textbf{8\_9}  \\
			\\
			\multirow{1}{*}{\textbf{Error ($10^{-2}$)} } & 1.87 & 3.07 & 3.64 & 2.20 & 2.99 & 4.85 & 3.84 & 3.01 & 3.42  \\
			\\
			\multirow{2}{*}{} & \textbf{0\_14} & \textbf{1\_13} & \textbf{2\_12} & \textbf{3\_11} & \textbf{4\_10} & \textbf{5\_9} &  &  &  &  &  \\
			\\
			& 2.65 & 6.94 & 3.92 & 4.25 & 2.63 & 3.37 &  &  &  &  & \\
			\\
			\\
			& \textbf{Q9} & \textbf{Q10} & \textbf{Q11} & \textbf{Q12} & \textbf{Q13} & \textbf{Q14} \\
			\\
			$T_1$, {\textmu}s & 36.4 & 76.4 & 55.9 & 91.0 & 26.6 & 42.3 \\
			\\
			$T_2$, {\textmu}s & 44.1 & 61.1 & 74.8 & 47.0 & 52.5 & 49.8 \\
			\\
			\textbf{Gate Error ($10^{-3}$)} & 2.11 & 1.54 & 0.79 & 2.79 & 1.97 & 0.67 \\
			\\
			\textbf{Readout Error ($10^{-2}$)} & 5.50 & 3.05 & 4.85 & 5.95 & 12.95 & 5.30 \\
			\\
			\multirow{1}{*}{\textbf{Multi-Qubit Gate (CX)}}&  \textbf{9\_10} & \textbf{10\_11}& \textbf{11\_12} & \textbf{12\_13} & \textbf{13\_14} &   \\
			\\
			\multirow{1}{*}{\textbf{Error ($10^{-2}$)} } & 3.61 & 3.24 & 8.32 & 10.84 & 6.63 &   \\
		\end{tabular}
	}
	\label{taberrors}
\end{table}

 Since the value of entanglement (\ref{geommeasurecat}) does not depend on the
parameters $\phi$ and $\lambda$, we set them equal to zero for $U_3$ gate. On the quantum computer, we prepared and measured the value of
entanglement between q[6] qubit and the remaining qubits. We measure q[6] qubit because it has the longest coherence time
$T_1$ and $T_2$, the smallest readout error and a rather small single-qubit gate error (see table~\ref{taberrors}).
These facts allow us to quite accurately measure the entanglement of the qubit with the remaining system.
We provide the measurements for different values of $\theta$, which changes in the range from 0 to $2\piup$ with the step $\piup/20$.
Thus, to obtain the value of entanglement of state with certain $\theta$, one should measure
the mean values of all the components of the spin corresponding to qubit q[6].
For each component of spin in a predefined pure quantum state, the quantum computer makes a specific
number of shots (in our case 1024). This allows us to obtain the probabilities to find this spin in states
$\vert 0\rangle$ and $\vert 1\rangle$. Then, we substitute these probabilities into equations (\ref{mvpauli}).
As a result, we obtain the mean values of the spin operator. Using these mean values in definition (\ref{geommeasure}),
we can obtain the value of entanglement between qubit q[6] and the remaining system.
Note that the more shots the quantum computer make, the smaller is the error due to the counting
statistics which is inversely proportional to the square root of the number of shots (in our case  $\sim 1/32$).

\begin{figure}[!t]
	\centering
\includegraphics[width=.33\textwidth]{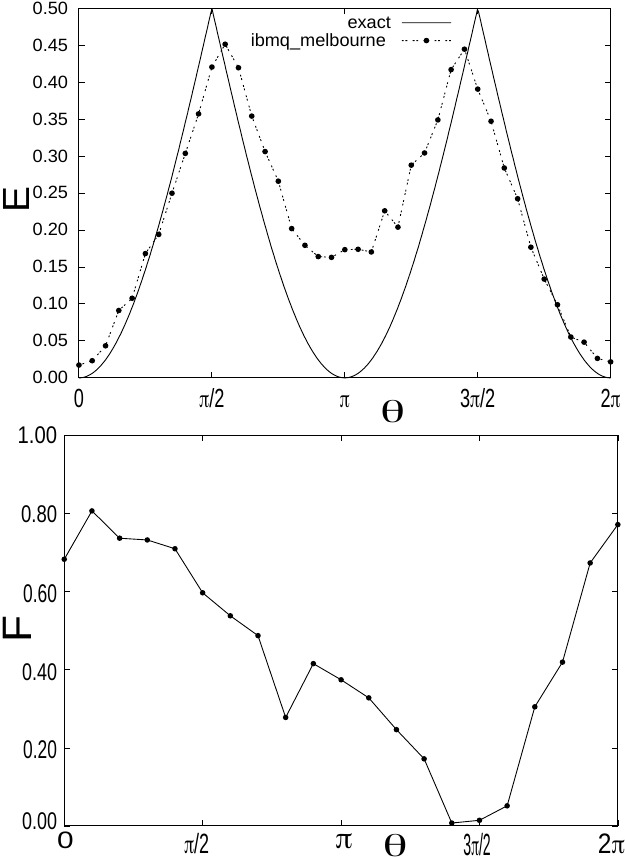}
\includegraphics[width=.32\textwidth]{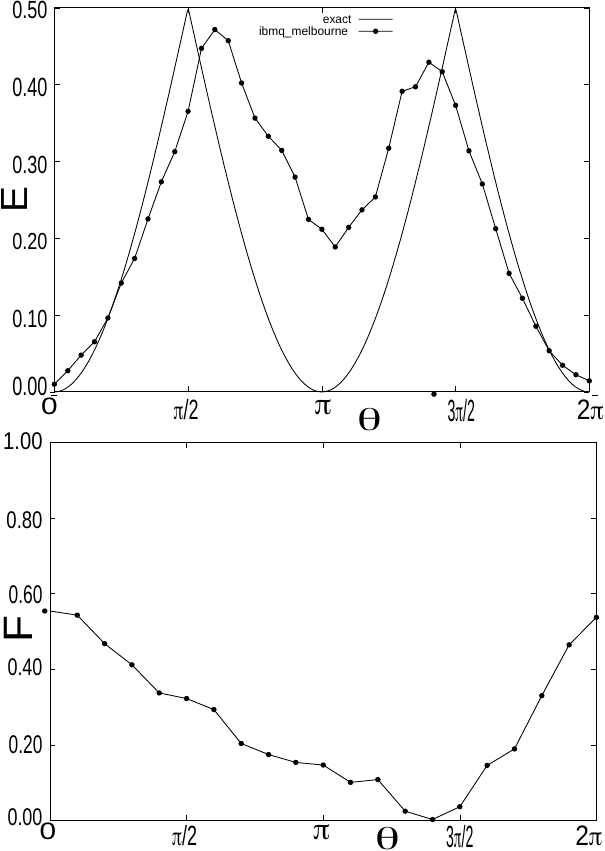}
\includegraphics[width=.33\textwidth]{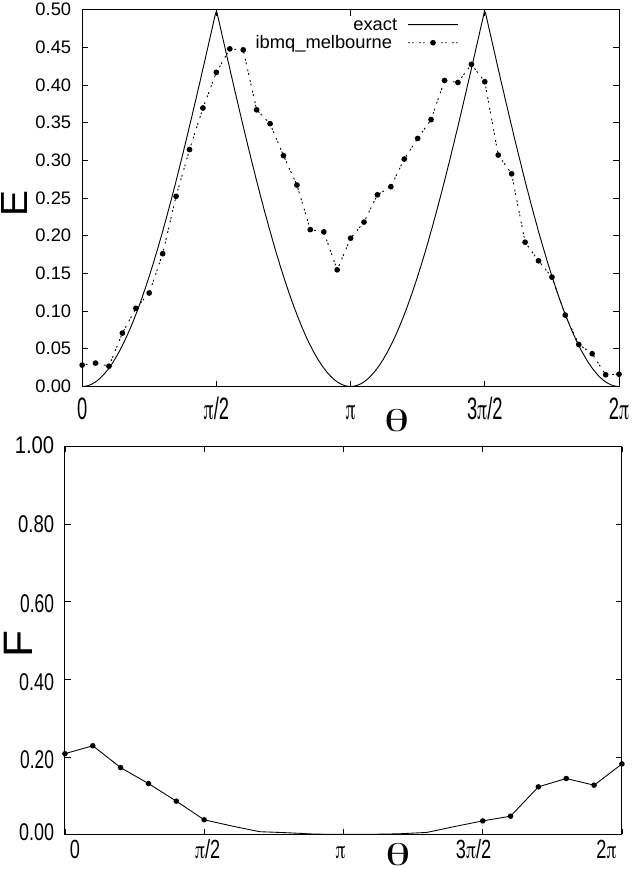}
\caption{Dependencies of the values of entanglement (upper figures) and fidelity (lower figures) of 5-qubit~(a), 10-qubit (b) and 15-qubit (c) Schr\"odinger cat state.
We compare experimental results (dots) with theoretical predictions (solid line) [equation (\ref{geommeasurecat})].}
\label{Hcatstate_5_10qubit}
\end{figure}

 The dependence of entanglement on parameter $\theta$ of q[6] qubit with the remaining system in the 5-, 10- and 15-qubit Schr\"odinger cat states is shown in figure~\ref{Hcatstate_5_10qubit}.
As we can see, the results are in good agreement with the theoretical prediction [see equation (\ref{geommeasurecat})]. However, the entanglement of states which are closer to the $\theta=\piup$
is somewhat different from the theoretical prediction. This  is caused by the relaxion time $T_1$ of each qubit from $\vert 1\rangle$ state to $\vert 0\rangle$ state (see table~\ref{taberrors}).
Deviations from theoretical predictions are also caused by the gate and readout errors.
To analyze this point more in detail, we measure the fidelity of the states which is studied. For this purpose, we calculate the folowing scalar product
\begin{eqnarray}
F=\vert\langle\psi_{\textrm{cat}}\vert\psi_{\textrm{cat}}^{\textrm{m}}\rangle\vert^2,
\label{fidelity}
\end{eqnarray}
where $\vert\psi_{\textrm{cat}}\rangle$ is the Schr\"odinger cat state defined by equation (\ref{Heisenbergstate}) and $\vert\psi_{\textrm{cat}}^{\textrm{m}}\rangle$ is the Schr\"odinger cat state prepared and
measured on the quantum computer. The dependence of fidelity on parameter $\theta$ in the 5-, 10- and 15-qubit Schr\"odinger cat states is shown in figure~\ref{Hcatstate_5_10qubit}.
As we can see, the fidelity decreases with distance from the $\vert 00\ldots 0\rangle$. Moreover, the fidelity depends on the number of qubits in the state. The greater is the number of qubits in the state,
the lower is the fidelity. However, we do not observe such a dependece on the number of qubits for entanglement. This is due to the fact that to determine the entanglement we measure only one qubit. Therefore, the readout error
accumulates from the measurement of this qubit. However, to determine the fidelity, we measure all the qubits of the state. In this case, the readout errors accumulate from all the qubits. This leads to a deterioration of fidelity of the states.
Finally, let us study the value of the entanglement of each spin with the rest of the system in the 15-qubit Schr\"odinger cat state for $\theta=\piup/2$.

\section{Entanglement of each qubit with the remaining system in the Schr\"o\-dinger cat state}

Recently, it was shown that 16-qubit \cite{wang2018} and 20-qubit ibmq \cite{mooney2019} devices can be fully entangled.
The authors of \cite{wang2018,mooney2019} prepared the graph state on quantum devices.
Using the negativity as a measure of entanglement, they showed that each pair of qubits in the graph states are entangled.

\begin{figure}[!t]
\centering
\includegraphics[scale=0.700, angle=0.0, clip]{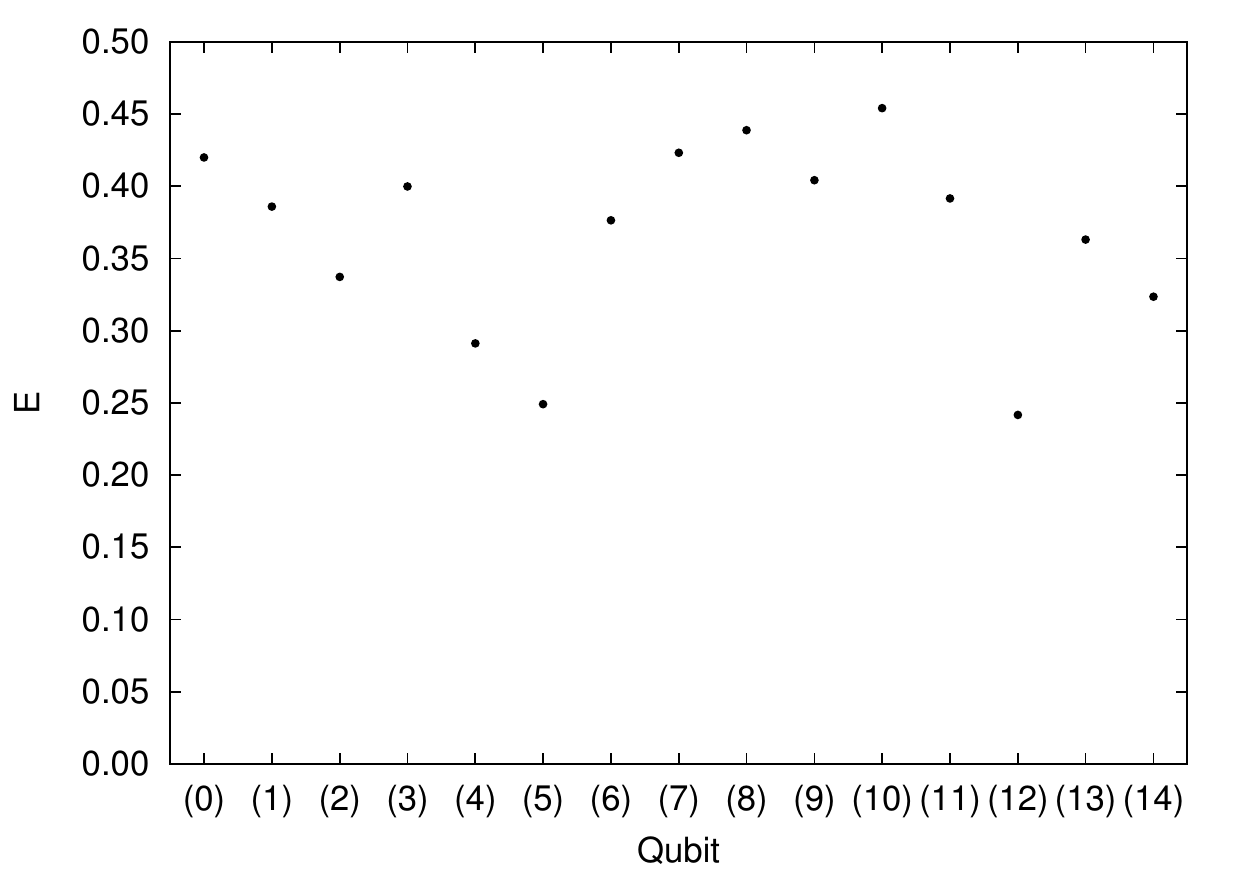}
\caption{Dependence of the value of entanglement of each qubit with other qubits in the Schr\"odinger cat state.
For all states, parameter $\theta=\piup/2$.}
\label{maxentallq}
\end{figure}

In this section, using our protocol, we make the investigation of entanglement of each qubit with the remaining system
in the state prepared on the ibmq-melbourne device. Namely, we prepare the maximum entangled
Shr\"odinger cat state (figure~\ref{ShCatstate}) with $\theta=\piup/2$. Further, we measure
the entanglement of each qubit with the remaining system.
Thus, measuring each qubit 1024 times  the quantum computer gives probabilities, which we substitute
in equtions (\ref{mvpauli}). Then, we put the obtained mean values
into definition (\ref{geommeasure}). Finally, we obtain the values of entanglement of each qubit with the remaining system.
The results are presented in figure~\ref{maxentallq}.
In all cases, theoretical prediction of entanglement is $E=1/2$. This is because theoretically
each qubit in the Schr\"odinger cat state with $\theta=\piup/2$ is maximum entangled with the remaining system.
However, the errors and coherence times of quantum device (see table~\ref{taberrors}) and systematic error of
the $U_3(\piup/2,0,0)$ operator spoil the experimental
results. This  leads to a decrease of entanglement in the system. For instance, in all states,
the results for the q[5] qubit have a worse agreement with the theoretical prediction than the results for the other
qubits (figure~\ref{maxentallq}). This is because this qubit has  shorter coherence times, the biggest single-qubit error
and a rather big readout error. The q[0] qubit is in the opposite situation. It has the smallest single-qubit error and
a rather small readout error.

\section{Conclusions}

We have used the protocol proposed in paper \cite{kuzmak2020} to determine the entanglement between a certain
qubit and the remaining system in the Schr\"odinger cat state prepared on the ibmq-melbourne quantum computer.
This protocol is based on the relation between the geometric measure of entanglement and the mean value of spin
corresponding to a certain qubit.
We have obtained dependencies of the values of entanglement
on state parameter of 5-, 10- and 15-qubit (figure~\ref{Hcatstate_5_10qubit}) Schr\"odinger cat states. Due to the fact that we have measured
only one qubit, the results are in good agreement with theoretical predictions. We have also studied the entanglement
of each qubit with the remaining system in the maximum entangled Schr\"odinger cat state (figure~\ref{maxentallq}).
Despite the errors that accumulate due to the preparation and measurement of the states, we observe a high level of entanglement.

\section{Acknowledgements}

This paper is dedicated to Oleg Derzhko and Andrij Shvaika on the occasion of their 60th
birth anniversaries.

We thank Prof. Andrij Rovenchak for useful comments.
This work was supported by Project FF-83F (No.~0119U002203) from the Ministry of Education and Science of Ukraine.

%
%

\ukrainianpart

\title{Приготування і дослідження заплутаності стану кота Шредінгера на квантовому комп'ютері ibmq-melbourne}
\author{А.Р. Кузьмак, В.М. Ткачук}
\address{
Кафедра теоретичної фізики, Львівський національний університет імені Івана Франка, \\вул. Драгоманова, 12
Львів, 79005, Україна 
}

\makeukrtitle

\begin{abstract}
\tolerance=3000%
Ми вивчаємо заплутаність одного кубіта з рештою кубітів системи, яка перебуває у стані ``кота Шредінгера'', який створений на квантовому комп'ютері ibmq-melbourne.
Протокол, який ми використовуємо для цього, ґрунтується на визначенні середнього значення спіну, що відповідає конкретному кубіту. Ми досліджуємо залежність заплутаності
від параметру стану ``кота Шредінгера'', який складається з різного числа кубітів. Також ми досліджуємо заплутаність кожного кубіту з рештою кубітів системи у
максимально заплутаному стані ``кота Шредінгера''.
\keywords стан ``кота Шредінгера'', заплутаність, квантові комп'ютери

\end{abstract}

\end{document}